\documentclass[comunication,twocolumn,showpacs,superscriptaddress]{revtex4}
\usepackage{amsthm,amssymb,amsfonts}
\usepackage{mathrsfs,color}
\usepackage{graphicx,hyperref}
\def\be{\begin{equation}}
\def\ee{\end{equation}}
\def\ba{\begin{eqnarray}}
\def\ea{\end{eqnarray}}

\usepackage{CJK}
\theoremstyle{plain}
\newtheorem{triM}{Property}

\begin{document}
%\begin{CJK*}{GBK}{song}
%\begin{CJK*}{UTF8}{gbsn}

\title{Coherent Destruction of Tunneling and Dark Floquet State}

%\author{Xiaobing Luo (¬ﬁ–°±¯)$^{1,3}$}
%\author{Liping Li (¿Ó¿˚∆Ω)$^{2,3}$}
%\author{Li You (”»¡¶)$^{3}$}
%\author{Biao Wu (Œ‚Ï≠)$^{4}$}
%\altaffiliation{Electronic address: wubiao@pku.edu.cn}
%\affiliation{$^{1}$Department of Physics, Jinggangshan University, Ji'an 343009, China}
%\affiliation{$^{2}$Information engineer school,
%Huanghe Science and Technology College, Zhengzhou 450006, China}
%\affiliation{$^{3}$Department of Physics, Tsinghua university,
%Beijing 100084, China}
%\affiliation{$^{4}$International Center for
%Quantum Materials, Peking University, Beijing 100871, China}

\author{Xiaobing Luo}
\affiliation{Department of Physics, Jinggangshan University,
Ji'an 343009, China}
\affiliation{Department of Physics, Tsinghua university,
Beijing 100084, China}
\author{Liping Li}
\affiliation{Information Engineer School,
Huanghe Science and Technology College, Zhengzhou 450006, China}
\affiliation{Department of Physics, Tsinghua university,
Beijing 100084, China}
\author{Li You}
\affiliation{Department of Physics, Tsinghua university,
Beijing 100084, China}
\author{Biao Wu}
\altaffiliation{Electronic address: wubiao@pku.edu.cn}
\affiliation{International Center for
Quantum Materials, Peking University, Beijing 100871, China}
\begin{abstract}
We study a  system of three coherently coupled states, where one state is shifted periodically against the other two. We discover such a system possesses a dark Floquet state at zero quasi-energy and always with negligible population at the intermediate state.
This dark Floquet state manifests itself dynamically in terms of the suppression of inter-state tunneling,    a phenomenon known as coherent destruction of tunneling.
We suggest to call it dark coherent destruction of tunneling (DCDT).
At high frequency limit for the periodic driving, this Floquet state reduces to the well-known dark state widely used for STIRAP.  Our results can be generalized to systems with more states and can be verified with easily implemented experiments within current technologies.

\pacs{42.65.Wi, 03.75.Lm, 33.80.Be, 42.82.Et}
\end{abstract}

\maketitle
Two-state and three-state models are the simplest quantum systems. Despite their simplicity, they often provide very good approximations to describe realistic physical systems and are capable of revealing a variety of fascinating quantum effects. The understanding of their ubiquitous features are nowadays being exploited for the manipulation and control of quantum states of small systems involving
single atoms, photons, or nano-devices~\cite{Brif,Salger,P.Hanggi}.
Coherent destruction of tunneling (two-state models) and dark state (three-state models) are two of the elegant prototype examples where deep understanding gained from quantum coherent effects in these simple systems are impacting the development of quantum technology in communication and computation.

Coherent destruction of tunneling (CDT) was discovered in
a periodically-driving double well system~\cite{Grossmann}. It describes
a fascinating phenomenon whereby coherent tunneling between two wells
(or the Rabi oscillation between two states) is turned off by an externally
enforced periodic level shift.  Its understanding is related to dynamical
localization~\cite{Dunlap}, which occurs at isolated degenerate points of
the quasienergies~\cite{Grossmann,Holthaus}. CDT has thus far generated significant interest, and has been theoretically extended into various forms~\cite{Creffield}-\cite{Kayanuma}.
It has also been observed experimentally in many physical systems :
including modulated optical coupler~\cite{Valle}, driven double-well potentials for single-particle tunneling~\cite{Kierig}, and a single electron spin
in diamond~\cite{Zhou}. More recently, it has also found application in tuning
the tunneling parameter of a Bose-Einstein condensate~\cite{Lignier,Eckardt}.

Dark state is often discussed in terms of a three-state system where two of them are coupled coherently to the intermediate state, as in the model system we study here. When all coupling fields are on resonance with their respective coupled pair of states, we can adopt the rotating wave approximation and change into a suitable interaction picture with all coupling coefficients becoming time independent.  In this case, there always exists a dark state, whose eigenenergy becomes uniformly zero, and the corresponding eigenvector contains no projection onto the intermediate state.
It is called dark as the intermediate state is an excited state capable of emitting photons. This type of dark state is also known as coherent population
trapping~\cite{Bergmann}, widely used in efficient population transfer through the stimulated Raman adiabatic passage (STIRAP) protocol. It has become the theoretical basis for several well-established implementations of quantum control and rudimentary quantum information processing gates.

In this Letter we report our surprising finding of an intimate relationship between dark state and coherent destruction of tunneling by studying a three-state system.  In this system,
two states are coherently coupled to an intermediate state and  one of the two states shifts
periodically against the other two.  We find that CDT also exists in this three-state system,
where the dynamical tunneling from one state to the other two is
suppressed by the periodic driving over a range of parameters. However,
this CDT for the three-state system has
its own distinct feature: it is related to a dark Floquet state, which
has zero quasi-energy and negligible population at the
the intermediate state. Quite interestingly, this dark Floquet state reduces to
the well-known dark state in a non-driving three-state
$\Lambda$-system\cite{Bergmann} at high-frequency limit.
Therefore, we call this CDT  \emph{dark coherent destruction of tunneling}(DCDT).
These results can be generalized to $N$-state system.
We also discuss a feasible experimental scheme where the visualization
of the DCDT  can be achieved readily.

\emph{Three-state system.} The driving three-state system  is described by the
Schr\"{o}dinger equation ($\hbar=1$)
%\begin{equation}
%i\frac{d}{dt}\pmatrix{c_1\cr c_2\cr c_3}=
%\pmatrix{\frac{A}{2}\sin(\omega t)&v&0\cr
%v&-\frac{A}{2}\sin(\omega t)&v\cr
%0&v&-\frac{A}{2}\sin(\omega t)\cr
%}\pmatrix{c_1\cr c_2\cr c_3}
%\end{equation}
\begin{eqnarray}
i\frac{dc_1}{dt}&=&~~\frac{A}{2}\sin(\omega
t)c_1+vc_{2},\nonumber\\
i\frac{dc_2}{dt}&=&-\frac{A}{2}\sin(\omega t)c_2+vc_{1}+vc_{3},\label{coupled}\\
i\frac{dc_3}{dt}&=&-\frac{A}{2}\sin(\omega t)c_3+vc_{2},\nonumber
\end{eqnarray}
where $c_1$, $c_2$, and $c_3$ are the amplitudes at three states $|1\rangle$, $|2\rangle$, and $|3\rangle$, respectively. $v$ is the coupling constant between the
neighboring states. Energy state $|1\rangle$ is shifted periodically against the other two
 with  driving strength $A$ and  frequency $\omega$.
The normalization condition $\Sigma_{j=1}^{3}|c_j|^2=1$ is
assumed.

\begin{figure}[htp]
\center
\includegraphics[width=8cm]{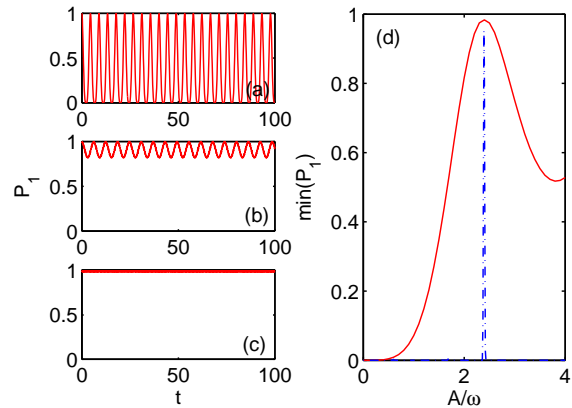}
\caption{(color online) The evolution of the probability
at state $|1\rangle$ $P_1=|c_1|^2$ for the system (\ref{coupled}) for various
driving conditions: (a) $A/\omega=0$; (b) $A/\omega=2.0$; (c) $A/\omega=2.4$.
(d) The minimum value of $P_1$ (solid line) as a function of driving parameter $A/\omega$.
The two-state results are plotted as a dash dotted line for comparison.
The initial condition is $\{c_1=1, c_2=0, c_3=0\}$. The other parameters
are $\omega=10,v=1$.} \label{fig1}
\end{figure}

To investigate the tunneling dynamics, we solve the
 time-dependent Schr\"{o}dinger equation
(\ref{coupled}) numerically with initial state $(1,0,0)^T$ .
The evolution of the probability distribution
$P_1=|c_1|^2$ is presented in Fig.~\ref{fig1} for three typical driving conditions.
For $A/\omega=0$ (Fig.~\ref{fig1} (a)), we see that $P_1$ oscillates
between zero and one, demonstrating no suppression of tunneling. For
$A/\omega=2.0$ (Fig.~\ref{fig1} (b)), the oscillations of $P_1$ are seen
limited between 0.8 and 1, showing suppression of tunneling. At
$A/\omega=2.4$ (Fig.~\ref{fig1} (c)), $P_1$ remains near unity,
signaling a complete suppression of tunneling between energy states. This
is  the quantum phenomenon well known as CDT~\cite{Grossmann}.

We emphasize that what we find here is not a simple re-discovery of CDT in a three-state system.
The CDT  in this three-state model has its own distinct feature: the results
shown in Fig. ~\ref{fig1} (b,c) indicate that the suppression of tunneling occurs
over a wide range of system parameters. This is in stark contrast to the CDT
in a two-state system~\cite{Grossmann}, which occurs
only at isolated points of parameters. The widening of the
suppression regime found in the driving
three-state system is more clearly demonstrated in Fig.~\ref{fig1}
(d), where the minimum value of $P_1$ is used to
measure the suppression of tunneling. When \textrm{min}($P_1$) is
not zero, the tunneling is suppressed as the population is not allowed to be
fully transferred from state $|1\rangle$ to the other two states. It is clear
from Fig.~\ref{fig1} (d) that the suppression occurs as long as $A/\omega\neq 0$.
For comparison, the results for the standard driven two-state system is plotted as dash dotted line in Fig.~\ref{fig1}
(d), where the extremely narrow peak width indicates that CDT occurs only at isolated points of parameters.
%({\color{red} more detailed description of Fig. 1(d) is needed; two-state results may
%need to be plotted in (d) for comparison.})

There exists a fundamental reason why the CDT occurs at isolated system parameters in
a two-state model, where the CDT is related to the degeneracy of
quasi-energy levels~\cite{Grossmann} and the degeneracy usually happens only
at isolated points. Therefore, the significant widening  of the suppression parameter range
that we see in Fig.~\ref{fig1} indicates that the CDT found here should have a
different origin. To investigate this origin, we turn to the Floquet theory for a
periodically-driving system. Similar to Bloch states for
systems with spatially periodic potentials, the modulated system
(\ref{coupled}) has Floquet states,
$(c_1,c_2,c_3)^T=(\tilde{c}_1,\tilde{c}_2,\tilde{c}_3)^T\exp(-i\varepsilon
t)$, where  $\varepsilon$ is  the quasi-energy and the  amplitudes
$(\tilde{c}_1,\tilde{c}_2,\tilde{c}_3)^T$ are periodic with modulation
period $T = 2\pi/\omega$.
%Any solution of the Schr\"{o}dinger equation
%(\ref{coupled}) can be decomposed into the linear superposition of all the Floquet states
%with time-independent expansion coefficients. Therefore, the dynamics of the
%system is dictated by Floquet states.

\begin{figure}[htp]
\center
\includegraphics[width=8cm]{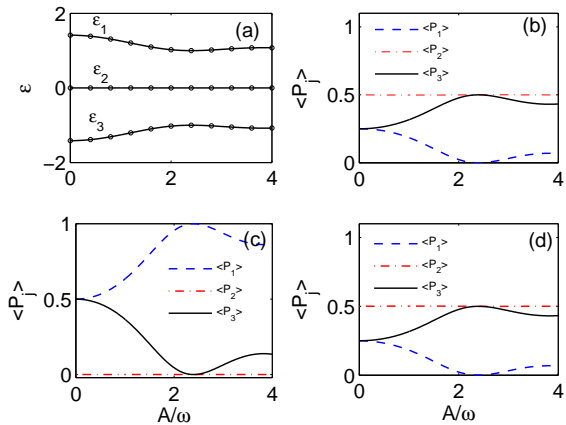}
\caption{(color online) (a) Quasienergies versus $A/\omega$.  Solid
lines are for numerical results obtained from the original model
(\ref{coupled}) and circles are for the approximation
results given by the effective model (\ref{eq:ceff}).
Time-averaged populations for the Floquet state in the quasi-energy
level (b) $\varepsilon_1$, (c) $\varepsilon_2$, and (d)
$\varepsilon_3$. The other parameters are $v=1,\omega=10$.}
\label{fig2}
\end{figure}

% The quantum dynamics of the modulated system (\ref{coupled}) can be
% clearly understood by analyzing the quasi-energies and corresponding Floquet states.
Our numerical results of the quasi-energies and Floquet states for
the modulated system (\ref{coupled}) are plotted in Fig. \ref{fig2}.
There are three Floquet states with quasi-energies $\varepsilon_1$, $\varepsilon_2$, and
$\varepsilon_3$. We immediately notice from Fig. \ref{fig2}(a) that the quasi-energy
$\varepsilon_2$ for the second Floquet state is always \emph{zero}  for all
values of $A/\omega$. We call this state {\it dark Floquet state} in analogy to
the well-known dark state. This dark Floquet state stands out not only for its zero
quasi-energy but also for its unique population distribution among energy states.
We display the time-averaged population probability $\langle
P_j\rangle==(\int_{0}^{T}dt |c_j|^2)/T$ for a given Floquet state
$(c_1,c_2,c_3)^T$ in Fig. \ref{fig2}(b-d). The Floquet state with $\langle
|P_j|^2\rangle> 0.5$ is generally regarded as a state localized at
the $j$-th energy state.  As seen in Fig. \ref{fig2}(c), the dark Floquet state  has almost
zero population at energy state $|2\rangle$ while the population at $|1\rangle$
$\langle P_1\rangle>0.5$ . In other words, the dark Floquet state
is localized at $|1\rangle$. The other two Floquet states
have identical population distribution. Since all their populations $\langle P_j\rangle \le 0.5$,
these two Floquet states are \emph{not} localized.

%%%%%%%%%%%%%%%%%%%%%%%%%
%%%%%%%%%%%%%%%%%%%%%%%%%%%

It is not difficult to see the suppression of tunneling seen in Fig.~\ref{fig1} is linked to
the existence of the dark Floquet state.  We expand the initial state in terms of the
Floquet states
\begin{equation}
(1,0,0)^T=b_1|\varepsilon_1\rangle+b_2|\varepsilon_2\rangle+b_3|\varepsilon_3\rangle\,.
\end{equation}
During the dynamical evolution, the expansion coefficient $b_i$ evolve as
$b_i\exp{(-i\varepsilon_i t)}$. In other words, $|b_i|$s are time independent.
We look at the case $A/\omega=2.4$, where $|\varepsilon_2\rangle$ has
population one at state $|1\rangle$ while the other two states have zero
population at $|1\rangle$. In this case, we have $|b_2|=1$ and $b_1=b_3=0$,
which corresponds to a complete suppression of tunneling from $|1\rangle$
to $|2\rangle$ and $|3\rangle$. For other values of $A/\omega$, similar arguments
can be made. This shows that the CDT observed in Fig.~\ref{fig1}
has a different origin: it is the consequence of dark Floquet states.
Therefore, we call it \emph{dark coherent destruction of tunneling} (DCDT).

Interestingly,  this dark Floquet state can be reduced to the well known
dark state  in a non-driven three-state $\Lambda$-system at high frequency limit.
Introducing the transformation  $c_m=a_m\exp[\pm iA\cos(\omega t)/(2\omega)]$
($+$ for $m=1$ and $-$ for $m=2,3$) and averaging out high frequency terms,
one can obtain a non-driven three-state system
\begin{eqnarray}
i\frac{d a_1}{dt}&=&vJ_0(A/\omega)a_2, \nonumber\\
i\frac{d a_2}{dt}&=&vJ_0(A/\omega)a_1+va_3,\label{eq:ceff} \\
i\frac{d a_3}{dt}&=&va_2, \nonumber
\end{eqnarray}
where $J_0(A/\omega)$ is the zeroth order Bessel function.
The famous dark state (also known as coherent trapped state) for Eq. (\ref{eq:ceff}) is
given by $(a_1,a_2,a_3)^T=\frac{1}{\mathcal{\sqrt{M}}}(-v,0,vJ_0(A/\omega))^T$, where
$\mathcal{M}=v^2+[vJ_0(A/\omega)]^2$.
This dark state corresponds to the dark Floquet state. Similarly, this dark state
is always localized at state $|1\rangle$ as $v>vJ_0(A/\omega)$ and has zero population
at state $|2\rangle$. This state is completely localized at state $|1\rangle$ when
$J_0(A/\omega)=0$.
%({\color{red} Is the first zero point of $J_0(A/\omega)$ close to 2.4?})
We have computed the the eigenvalues  of model (\ref{eq:ceff})
and compared them (circles) with the quasi-energies (black solid lines)
in Fig. \ref{fig2} (a). The agreement is almost perfect.

\emph{Generalization to $N$-state system.}
Our analysis above is given for a three-state system and the original CDT was found
in a two-state system.
These results can be generalized to an  $N$-state system, where one state
is shifted periodically against all the other states. The equations of motion are
\begin{eqnarray}
i\frac{dc_j}{dt}&=&v(c_{j-1}+c_{j+1})+E_j(t)c_j,\label{coupledN}\\
E_1(t)&=&\frac{A}{2}\sin(\omega t),~~E_{j\neq 1}(t)=-\frac{A}{2}\sin(\omega
t),\nonumber
\end{eqnarray}
where $c_{j\leq 0}=c_{j>N}=0$.

\begin{figure}[htp]
\center
\includegraphics[width=8cm]{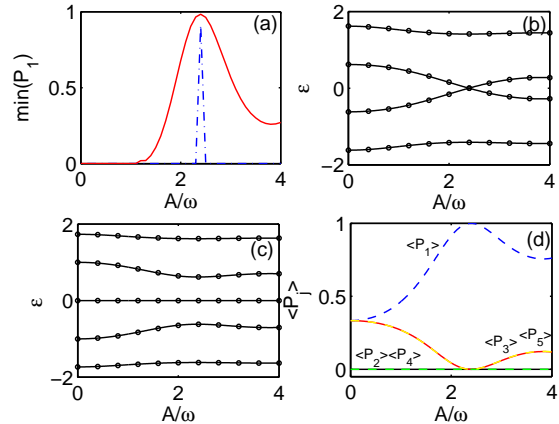}
\caption{(color online) (a) The minimum value of $P_1$ as a function
of $A/\omega$ for $N=4$ (dash dotted line) and $N=5$ (solid
line). The initial conditions are $c_1(0)=1, c_j(0)=0 (j\neq 1)$.
Quasi-energies versus $A/\omega$ for (b) $N=4$ and (c) $N=5$. Solid
lines are for numerical results obtained from the model
(\ref{coupledN}) and circles are for the approximation
results given by the effective model (\ref{effective}). (d) The
time-averaged probability distribution of the Floquet state
corresponding to $\varepsilon=0$ in Fig. \ref{fig3} (c). The other
parameters are $v=1,\omega=10$.} \label{fig3}
\end{figure}

The quantum dynamics of the driven $N$-state systems is investigated
by direct integration of the time-dependent Schr\"{o}dinger equation
(\ref{coupledN}) with the state initially prepared on the state $|1\rangle$.
The CDT is found to exist. The minimum value of $P_1=|c_1|^2$ as a
function of $A/\omega$ is presented in Fig. \ref{fig3} (a) for $N=4$ and $N=5$.
When $N=4$, the CDT occurs at an isolated point of
parameters (dash dotted line in Fig. \ref{fig3} (a)),  where two of the four
quasi-energy levels become degenerate (Fig. \ref{fig3} (b)). This is exactly
the same as in the two-state system. When $N=5$,  the parameter
range where CDT occurs  is extended substantially (solid line in Fig. \ref{fig3} (a))
as in the three-state model. Furthermore, this five-state system
also has a dark Floquet state: as seen in Fig. \ref{fig3} (c), one of  the  quasi-energies
always equals to zero. This dark Floquet state has negligible
population at all of even $j$-th states (Fig. \ref{fig3} (d)).

These numerical results with $N=4,5$, together with the known results for $N=2,3$,
clearly suggest that ({\it i}) the dark state and the associated DCDT exist in odd $N$-state systems;  ({\it ii}) the original CDT, which occurs at
isolated parameter points, exists in all even-$N$-state system. This general conclusion
can be proved analytically at high frequency limit.

Following the procedure used in the three-state system, we introduce the
transformation $c_1=a_1\exp[-i\int A\sin(\omega t)/2dt]$,
$c_{j\neq 1}=a_j\exp[i\int A\sin(\omega t)/2dt]$, where $a_j(z)$
are slowly varying functions. Using the expansion $\exp[\pm
iA\cos(\omega t)/\omega]=\sum_k(\pm i)^k J_k( A/\omega)\exp(\pm ik
\omega t)$ in terms of Bessel functions and neglecting all orders
except $k=0$ for high frequency limit,  we can reduce the coupled equations
(\ref{coupledN})  to a non-driven model
\begin{eqnarray}
i\frac{d \textbf{a}}{dt}=\bar{H}\textbf{a},
 \label{effective}
\end{eqnarray}
where $\textbf{a}=(a_1,a_2,...,a_N)^T$. The matrix $\bar{H}$ is tridiagonal  with
$\bar{H}_{12}=\bar{H}_{21}=v_{\rm{eff}}=vJ_0(A/\omega)$,
$\bar{H}_{n,n+1}=\bar{H}_{n+1,n}=v$. The
effective coupling constant $v_{\rm{eff}}$ between state $|1\rangle$ and state $|2\rangle$ is
tunable with the driving parameters.

%We search the stationary solution of the effective model
%(\ref{effective}) in the form $\textbf{a}=e^{-i\lambda
%t}\textbf{w}$, where $\lambda$ is the real eigenvalue, and the
%vector $\textbf{w}=(w_1,w_2,...,w_N)^T$ follows the linear
%eigenvalue equation
%\begin{eqnarray}
%\bar{H}\textbf{w}=\lambda\textbf{w}.
%\label{eigenvalue}
%\end{eqnarray}
%It is clear that the eigenvectors and eigenvalues of the above
%time-independent
%equation (\ref{eigenvalue})  correspond to the Floquet states and quasi-energies of
%driven model (\ref{coupled}).
%We have also computed the eigenvalues of Eq. (\ref{eigenvalue}) for $N=4$ and $N=5$,
%which are plotted as red dashed lines
%in Fig. \ref{fig3}(b-c). The analytical
%results for the quasi-energies from model (\ref{eigenvalue}) are in
%good agreement with the exactly numerical results from the original
%model (\ref{coupledN}).

The eigenvalues and eigenvectors of the tridiagonal $N\times N$
matrix $\bar{H}$ enjoy some very interesting properties, whose
rigorous proofs can be found in Supplemental Material.

(a) \emph{When $N$ is even, for any nonzero $v_{\rm{eff}}$ and $v$,  all
the eigenvalues of the matrix $\bar{H}$ are nonzero while two of
the eigenvalues are zero for $v_{\rm{eff}}=0$.}

\emph{Remark}: This means that when $N$ is even,
two quasi-energy levels of the driven model (\ref{coupledN}) are
degenerate at isolated points where $v_{\rm{eff}}=vJ_0(A/\omega)=0$.
The CDT occurs at these isolated points.

(b) \emph{When $N$ is odd,  one
and only one eigenvalue of $\bar{H}$ always equals to zero and,
for the corresponding eigenvector $(w_1,w_2,...,w_N)^T$ of $\bar{H}$,
the inequality $|w_1|^2>0.5$ holds for a finite range of parameters; for
any other eigenvector $(w_1,w_2,...,w_N)^T$ of $\bar{H}$ , one has
$|w_j|^2\leq 0.5$.}

\emph{Remark}: When $N$ is odd, the system always has one and only
one dark Floquet state, which is localized over a finite range of parameters.
Correspondingly, DCDT occurs over a finite range of parameters.

\emph{Experimental observation.}  By mapping the temporal evolution of
quantum systems into the spatial propagations of light waves, the
engineered waveguides have provided an ideal platform to investigate
a wide variety of coherent quantum effects\cite{Garanovich,Longhi}.
The phenomenon of DCDT can also be observed with this kind of waveguide
system. The discrete model (\ref{coupledN}) can be simulated by the light propagation in
an array of $N$  waveguides placed closely and with equal spacing.
Periodic driving is realized by the harmonic modulation of the refractive
index of the waveguides along the propagation direction\cite{Szameit,Szameit2}.
For our system, the periodic modulation of the first waveguide has a phase difference
of $\pi$ against the modulations for all other $N-1$ waveguides.  When $N$
is odd, the DCDT can be readily observed with current experimental
capacity\cite{Szameit,Szameit2}.

%The photonic lattice also enables a
%direct visualization of the dark Floquet state. We consider a
%three-channel waveguides with out-of-phase sinusoidal modulations of
%the refractive index between the first waveguide and the other two
%waveguides along the propagation direction. In the meantime, the
%amplitude of the sinusoidal modulation is slowly increased from zero
%to a constant within certain cycles. If the two outer waveguides are
%excited in their fundamental modes at the input plane, which
%corresponds to that the system is initially prepared in the dark
%state with equal population in the two outer waveguides ($
%P_1=P_3=0.5$ at $A/\omega=0$, see Fig.
%\ref{fig2} (c)), the system will track the dark Floquet state
%adiabatically. Therefore, the light transfer between two outer
%waveguides with negligible excitation of the central waveguide can
%be observed.

%\begin{eqnarray}
%\bar{P}_1=|b_1|^2=\Big[\frac{\Omega^2}{J^2+\Omega^2}+\frac{J^2}{J^2+\Omega^2}\cos(\sqrt{J^2+\Omega^2}z)\Big]^2.\label{eq:c8}
%\end{eqnarray}

In summary, we find that the CDT also happens
in a three-state quantum system, where one energy state
is shifted periodically against the other two states. We call this type of CDT
dark coherent destruction of tunneling (DCDT) as
it  is related to the existence of a dark Floquet state in the
three-state system. The dark Floquet state has zero
quasi-energy and negligible population at the intermediate state. It
reduces to the well known dark state of a non-driven three-state
system.  These results can be  generalized to a periodically
driven $N$-state system.   We have
also pointed out that observation of  DCDT  is well within the capacity of current
experiments.

 This work is funded by the NSF of
China (10965001, 11165009),  the NSF of Jiangxi Province (2010GQW0033),
 the Jiangxi Young Scientists Training Plan
(20112BCB23024), the financial support provided by the
Key Subject of Atomic and Molecular Physics in Jiangxi Province and
Zhengzhou MBOST (20120409). B. W. is supported by the NBRP of
China (2012CB921300,2013CB921900), the NSF of
China (11274024), the RFDP of China
(20110001110091). L. Y. is supported by the NSF of
China (11004116) and by
MOST 2013CB922000 of the National Key Basic Research Program of China.

\section*{Supplemental Material}
The $N\times N$ tridiagonal matrix $\bar{H}$ has the following non-zero matrix elements:
$\bar{H}_{12}=\bar{H}_{21}=v_{\rm{eff}}$, $\bar{H}_{n,n+1}=\bar{H}_{n+1,n}=v$ for $n=1, 2,\ldots, N-1$. $v_{\rm{eff}}$ is tunable and $v\neq 0$ is fixed.  The eigenvalues
and eigenvectors of $\bar{H}$ have the following properties. \\

\begin{triM}
When $N$ is odd, one
and only one eigenvalue of $\bar{H}$ always equals to zero.
\end{triM}
\begin{proof}
Let $\lambda_1,\lambda_2,\ldots,\lambda_N$ be all the
eigenvalues of  matrix $\bar{H}$, then
$D_N=\textrm{det}(\bar{H})=\lambda_1\lambda_2\cdots\lambda_N$.
It is easy to verify that $D_2=-v_{\rm{eff}}^2$, $D_1=D_3=0$, and
the relation $D_N=-v^2D_{N-2}
(N\geq 3)$. Therefore,  one has $D_{2k-1}=0$ and
$D_{2k}=(-1)^kv^{2k-2}v_{\rm{eff}}^2$ ($k=1,2,3,\ldots$).
When $N$ is odd, $D_{N}=\lambda_1\lambda_2\cdots\lambda_N= 0$, which
means that at least one eigenvalue equals to zero regardless the values of
$v_{\rm{eff}}$ and $v$. For the zero eigenvalue, the eigen-equation is
$\bar{H}\textbf{w}=0$, where $\textbf{w}=(w_1,w_2,\ldots,w_N)^T$. When expanded, the equation is turned into the following equations: $v_{\rm{eff}}w_2=0, v_{\rm{eff}}w_1+vw_3=0, vw_{j-1}+vw_{j+1}=0~~(j=3,4,\ldots,N-1), vw_{N-1}=0.$
There is only one non-trivial solution. For $v_{\rm {eff}}\neq 0$,  it  is given by
\begin{equation}
\textbf{w}=\frac{1}{\mathcal{\sqrt{M}}}(w^0_1,w^0_2,\ldots,w^0_N)^T
\label{ev}
\end{equation}
with
$w^0_1=(-1)^{(N-1)/2}v/v_{\rm eff}$, $w^0_{2k}=0$, and
$w^0_{2k+1}=(-1)^{(N-2k-1)/2}$ where $k=1,2,\ldots,(N-1)/2$ and
$\mathcal{M}=\sum_{j=1}^N |w_j^0|^2$ is the normalization factor.
For $v_{\rm eff}= 0$, the solution is $\textbf{w}=(1,0,\ldots,0)^T$.
This shows matrix $\bar{H}$ has one and only  one eigenvalue equal to zero.\\
\end{proof}

\begin{triM}
When $N$ is even, for any nonzero $v_{\rm{eff}}$,  all
the eigenvalues of the matrix $\bar{H}$ are nonzero while two of
the eigenvalues are zero for $v_{\rm{eff}}=0$.
\end{triM}

\begin{proof}
If $v_{\rm{eff}}\neq 0$, when $N$ is even, then
$D_{N}=(-1)^{N/2}v^{N-2}v_{\rm{eff}}^2=\lambda_1\lambda_2\cdot\cdot\cdot\lambda_N\neq 0$. Thus all
the eigenvalues of the matrix $\bar{H}$ are nonzero. When
$v_{\rm{eff}}=0$, it is obvious that
$D_{N}=\lambda_1\lambda_2\cdot\cdot\cdot\lambda_N=0$. There are zero eigenvalues.
With $v_{\rm{eff}}=0$,  the tridiagonal matrix of
$\bar{H}$ is divided into two uncoupled subspaces, i.e., $\bar{H}=0\oplus F$,
where $F$ is a tridiagonal matrix with nonzero elements
$F_{n,n+1}=F_{n+1,n}=v$.  It is clear that $F$ possesses Property 1 and has only  one
zero eigenvalue. $\bar{H}$ thus has two zero eigenvalues.
\end{proof}

\begin{triM}
If $\lambda$ is an eigenvalue of $\bar{H}$ with
eigenvector $(w_1,w_2,...,w_N)^T$, then $-\lambda$ is an eigenvalue
of $\bar{H}$ with the corresponding eigenvector
$(w'_1,w'_2,...,w'_N)^T$ where $ w'_j=(-1)^j w_j$.
\end{triM}

\begin{proof}
The eigenvalue equation
 $\bar{H}\textbf{w}=\lambda\textbf{w}$ can be written in the form
 $\Sigma_{j=1}^{N}\bar{H}_{ij}w_{j}=\lambda w_i$, where $\bar{H}_{ij}=0$ when
 $|i-j|=0$ and $|i-j|\geq 2$. Multiplying by $(-1)^{i-1}\lambda$, we obtain $\Sigma_{j=1}^{N}\bar{H}_{ij}(-1)^jw_{j}=-\lambda (-1)^i
 w_i$ and have the proof.
\end{proof}

\begin{triM}
When $N$ is odd, for the eigenvector
$(w_1,w_2,...,w_N)^T$ of $\bar{H}$ belonging to $\lambda=0$, the
inequality $|w_1|^2>1/2$ holds for a finite range of parameters; For
an eigenvector $(w_1,w_2,...,w_N)^T$ of $\bar{H}$ belonging
to $\lambda\neq 0$, one has that $|w_j|^2\leq 1/2$, whether $N$ is
odd or even.
\end{triM}

\begin{proof}
According to Eq.(\ref{ev}), it is clear that
the inequality $|w_1|^2>0.5$ is valid only when
$(v/v_{\rm{eff}})^2>(N-1)/2$. \\
When $\lambda\neq 0$, the two eigenvalues $\lambda$ and $-\lambda$
are distinct and the corresponding eigenvectors are
orthogonal to each other.  With Property 3,  one has
$\Sigma_{k=1}^{(N+1)/2}|w_{2k-1}|^2=\Sigma_{k=1}^{(N-1)/2}|w_{2k}|^2$
when $N$ is odd;
$\Sigma_{k=1}^{N/2}|w_{2k-1}|^2=\Sigma_{k=1}^{N/2}|w_{2k}|^2$ when
$N$ is even. With the normalization condition
$\Sigma_{j=1}^{N}|w_j|^2=1$, we immediately obtain that $|w_j|^2\leq
0.5$.
\end{proof}

%\end{CJK*}

\end{document}